\documentclass[10pt,conference]{IEEEtran}
\IEEEoverridecommandlockouts
\usepackage{cite}
\usepackage{amsmath,amssymb,amsfonts}
\usepackage{algorithmic}
\usepackage{graphicx}
\usepackage{textcomp}
\usepackage[table]{xcolor}
\usepackage{array}
\usepackage{paralist}
\usepackage{caption}
\usepackage{tikz,pgfplots,balance}
 \def\BibTeX{{\rm B\kern-.05em{\sc i\kern-.025em b}\kern-.08em
    T\kern-.1667em\lower.7ex\hbox{E}\kern-.125emX}}

\usepackage[utf8]{inputenc} 
\usepackage[T1]{fontenc}    
\usepackage{hyperref}       
\usepackage{url}            
\usepackage{booktabs}       
\usepackage{amsfonts}       
\usepackage{nicefrac}       
\usepackage{microtype}      
\usepackage{xcolor}         
\usepackage[nolist]{acronym}
\usepackage{algorithm,algorithmic}
\usepackage[font=scriptsize]{caption}
\usepackage{bm}
\usepackage[utf8]{inputenc} 

\usepackage{amsmath,dsfont,bbm,epsfig,amssymb,amsfonts,amstext,verbatim,amsopn,cite,subfigure}
\usepackage{multirow,multicol,lipsum,xfrac}
\usepackage{amsthm}
\usepackage{mathtools,amsthm}
\usepackage{url}
\usepackage{amsfonts}
\usepackage{tikz,pgfplots}
\usepackage[nolist]{acronym}
\usepgfplotslibrary{fillbetween}

\newcommand{\setR}{\mathbbmss{R}}

\newcommand{\setC}{\mathbbmss{C}}

\newcommand{\Ex}[2]{ \mathbbm{E}_{#2} \left[ #1 \right] }

\newcommand{\her}{\mathsf{H}}

\newcommand{\argmin}{\mathop{\mathrm{argmin}}}

\newcommand{\man}{\mathcal{N}}

\newcommand{\mac}{\mathcal{C}}

\newcommand{\bxx}{\mathbf{x}}

\newcommand{\bww}{\mathbf{w}}
\newcommand{\buu}{\mathbf{u}}

\newcommand{\bvv}{\mathbf{v}}
\newcommand{\byy}{\mathbf{y}}

\newcommand{\bh}{{\mathbf{h}}}

\newcommand{\bmu}{\boldsymbol{\mu}}

\newcommand{\set}[1]{\left\lbrace#1\right\rbrace}

\newcommand{\brc}[1]{\left( #1 \right) }

\newcommand{\dbc}[1]{\left[ #1 \right] }

\newcommand{\derivative}[1]{\frac{\partial}{\partial #1}}

\newcommand{\bzz}{{\mathbf{z}}}
\newcommand{\baa}{{\mathbf{a}}}

\newcommand{\trp}{\mathsf{T}}

\newcommand{\mI}{\mathbf{I}}

\newcommand{\mG}{\mathbf{G}}

\newcommand{\mSigma}{\boldsymbol{\Sigma}}

\newcommand{\mB}{\mathbf{B}}

\newcommand{\mH}{\mathbf{H}}

\newcommand{\MMSE}{\mathsf{MMSE}}

\newcommand{\ML}{\mathsf{ML}}

\newcommand{\diag}{\mathsf{diag}}

\newcommand{\abs}[1]{\lvert #1 \rvert}

\newtheoremstyle{mystyle}
{}
{}
{\it}
{}
{\bfseries}
{:}
{ }
{}

\theoremstyle{mystyle}

\newtheorem{definition}{Definition}

\newcounter{bar}

\begin{document}

\title{Global Unknown Estimation: A Statistical Framework for Wireless Distributed Learning}

\author{\IEEEauthorblockN{Yicheng Qu}
\IEEEauthorblockA{\textit{University of Toronto} \\
eason.qu@mail.utoronto.ca}
\and
\IEEEauthorblockN{Ali Bereyhi}
\IEEEauthorblockA{\textit{University of Toronto} \\
ali.bereyhi@utoronto.ca}
\and
\IEEEauthorblockN{Ben Liang}
\IEEEauthorblockA{\textit{University of Toronto} \\
liang@ece.utoronto.ca}

\thanks{This work was supported in part by the Natural Sciences and Engineering Research Council of
Canada.}

}

\maketitle

\begin{acronym}
	\acro{mimo}[MIMO]{multiple-input multiple-output}
	\acro{simo}[SIMO]{single-input multiple-output}
	\acro{csi}[CSI]{channel state information}
	\acro{awgn}[AWGN]{additive white Gaussian noise}
	\acro{iid}[i.i.d.]{independent and identically distributed}
	\acro{uts}[UTs]{user terminals}
	\acro{ps}[PS]{parameter server}
	\acro{irs}[IRS]{intelligent reflecting surface}
	\acro{tas}[TAS]{transmit antenna selection}
	\acro{glse}[GLSE]{generalized least square error}
	\acro{rhs}[r.h.s.]{right hand side}
	\acro{lhs}[l.h.s.]{left hand side}
	\acro{wrt}[w.r.t.]{with respect to}
	\acro{rs}[RS]{replica symmetry}
	\acro{mac}[MAC]{multiple access channel}
	\acro{np}[NP]{non-deterministic polynomial-time}
	\acro{papr}[PAPR]{peak-to-average power ratio}
	\acro{rzf}[RZF]{regularized zero forcing}
	\acro{snr}[SNR]{signal-to-noise ratio}
	\acro{sinr}[SINR]{signal-to-interference-and-noise ratio}
	\acro{svd}[SVD]{singular value decomposition}
	\acro{mf}[MF]{matched filtering}
	\acro{gamp}[GAMP]{generalized AMP}
	\acro{amp}[AMP]{approximate message passing}
	\acro{vamp}[VAMP]{vector AMP}
	\acro{map}[MAP]{maximum-a-posterior}
	\acro{ml}[ML]{maximum likelihood}
	\acro{mse}[MSE]{mean squared error}
	\acro{mmse}[MMSE]{minimum mean squared error}
	\acro{ap}[AP]{average power}
	\acro{ldgm}[LDGM]{low density generator matrix}
	\acro{tdd}[TDD]{time division duplexing}
	\acro{rss}[RSS]{residual sum of squares}
	\acro{rls}[RLS]{regularized least-squares}
	\acro{ls}[LS]{least-squares}
	\acro{erp}[ERP]{encryption redundancy parameter}
	\acro{zf}[ZF]{zero forcing}
	\acro{ta}[TA]{transmit-array}
	\acro{ofdm}[OFDM]{orthogonal frequency division multiplexing}
	\acro{dc}[DC]{difference of convex}
	\acro{bcd}[BCD]{block coordinate descent}
	\acro{mm}[MM]{majorization-maximization}
	\acro{bs}[BS]{base-station}
	\acro{aircomp}[AirComp]{over-the-air-computation}
	\acro{ULA}[ULA]{uniform linear array}
	\acro{ota-fl}[OTA-FL]{over-the-air distributed learning}
	\acro{los}[LoS]{line-of-sight}
	\acro{nlos}[NLoS]{non-line-of-sight}
	\acro{aoa}[AoA]{angle of arrival}
	\acro{nn}[NN]{neural network}
	\acro{cnn}[CNN]{convolutional neural network}
	\acro{sgd}[SGD]{stochastic gradient descent}
	\acro{aircomp}[AirComp]{over-the-air computation}
	\acro{lmi}[LMI]{linear matrix inequalities}
	\acro{qcqp}[QCQP]{quadratically constrained quadratic programming}
	\acro{lln}[LLN]{law of large numbers}
	\acro{clt}[CLT]{central limit theorem}
	\acro{gs}[GS]{gradient sparsification}
	\acro{ota-f}[OTA-F]{over-the-air fair}
	\acro{bc}[BC]{broadcast channel}
	\acro{pdf}[PDF]{probability density function}
	\acro{mvu}[MVU]{minimum variance unbiased}
    \acro{crlb}[CRLB]{Cram\'er--Rao lower bound}
    \acro{kl}[KL]{Kullback-Leibler}
    \acro{gue}[GUE]{global unknown estimation}
    \acro{slsqp}[SLSQP]{sequential least squares quadratic programming}
\end{acronym}

\begin{abstract}
Over-the-air computation (AirComp) is widely used for model aggregation in wireless distributed learning. Although it enhances communication efficiency, we believe the AirComp aggregation has limited effectiveness due to the difference between its target problem and that of distributed learning. In this paper, we develop a rigorous formulation for \textit{optimal} model aggregation in wireless distributed learning. Using this formulation, we show that AirComp aggregation generally assumes a \textit{mismatched} statistical model for local parameters. We then propose a statistical framework for model aggregation, called global unknown estimation (GUE). It captures the statistical relation between the local and global model parameters, allowing to interpret model aggregation as an inference task. We validate the efficiency of GUE through numerical experiments. Our results show that, in the low SNR regime, GUE can reduce the required power for model aggregation by approximately 15 dB compared to AirComp aggregation. Remarkably, this gain is obtained without additional computational overhead. 
\end{abstract}

%
\section{Introduction}
\par In distributed learning, multiple clients collaboratively train a model, potentially with the help of a central server, without sharing their local datasets. A well-known approach is federated learning \cite{mcmahan2017FL},
where the training is carried out through multiple rounds of communication: in each round, clients train local models on their own datasets and share the models with the server. The server \textit{aggregates} these models into a global model and broadcasts it to the clients to initiate the next round of training. This paradigm has gained significant importance in modern large-scale and privacy-sensitive applications~\cite{Arbaoui2024FLSurvey},
as it allows computation to be pushed closer to where data are stored. 
Under perfect communication links, the model transmission and aggregation can be readily performed~\cite{Li2020FedAvgConvergence}.

\par In practical wireless networks, however, the clients are connected through noisy communication links with imperfect channel conditions. This introduces a fundamental bottleneck 
\cite{Gafni2022FL_SPperspective}. To address this challenge, various approaches for efficient model aggregation in these systems have been proposed. Among these proposals, 
\ac{aircomp} has been widely accepted as the state-of-the-art \cite{Yang2020AirCompFL}. 
In this approach, instead of recovering transmitted signals separately and then aggregating them afterwards, the server directly estimates the desired aggregation by leveraging the superposition of waveforms over the wireless channel~\cite{CompMAC}. 
This enables simultaneous transmissions over the same channel, enhancing the communication efficiency.

\par From the statistical viewpoint, \ac{aircomp} implicitly assumes that realizing the predefined aggregation policy that is deployed in an error-free network, \emph{e.g.}, weighted averaging, is the optimal approach to aggregate a global update from the noisy received signals. 
Although this approach leads to a valid aggregation, its optimality is not guaranteed. In fact, using \ac{aircomp}, the global update is determined by minimizing the \textit{computation error} between the aggregated update and a predefined function of local models. How explicitly this computation error affects the learning performance, however, remains unclear. 
\par Several studies characterize the efficiency of \ac{aircomp} aggregation by connecting its computation error to learning metrics. The work in \cite{FedAvg-convergence-analysis} shows that the computation error determines an upper bound on the optimality gap of the learning algorithm. 
The studies in \cite{channel-adaptive-FL} and \cite{OTA-FL-heterogenous} extend this analysis to settings with imperfect \ac{csi} and heterogeneous data, respectively, and use this connection to propose adaptive aggregation schemes. 
The study in \cite{Seyed2024WeightedAgg} considers a joint design in settings with both imperfect \ac{csi} and heterogeneous data and develops an alternative aggregation scheme that directly minimizes an upper bound on the optimality gap. 
This idea is further extended to digital aggregation in \cite{Azimi2024Lattice} by implementing a coded \ac{aircomp} approach via lattice coding. All existing studies commonly focus on optimality bounds in terms of computation error and assume that minimizing this error \textit{directly} enhances learning performance.
\par In this study, we challenge this assumption and show that there is no strong connection between these two metrics. Intuitively, the goal of model aggregation is not to realize a predefined function, but rather to provide an \textit{efficient estimate} of the global update from the received signals. This describes a distributed inference task, in which a global unknown is to be estimated from a set of distributed observations; see for instance \cite{lower-bound-est-view,OTA-estimation}. Despite its intuitive connection, this framework has not been adapted to distributed learning. Our study aims to establish a novel framework that exploits the connection between these two concepts for efficient model aggregation in wireless networks. 

\subsection{Contributions} 
This work proposes a new framework, called \ac{gue}, for model aggregation in wireless distributed learning systems. \ac{gue} deviates from the conventional \ac{aircomp} approach and interprets the aggregation task as a distributed inference problem. This enables us to define the notion of \textit{optimal} aggregation and specify it under a postulated prior distribution for local parameters. Our experiments demonstrate that \ac{gue} can significantly enhance the power efficiency of model aggregation by shifting its focus from \textit{computation error minimization} to the \textit{optimal estimation of global model parameter}. 
\par More precisely, the main contributions of this work are as follows:
\begin{itemize}
    \item 
    We formulate model aggregation as a statistical inference problem and define the notion of \textit{optimal aggregation}. By fitting \ac{aircomp} aggregation into this framework, we show that \ac{aircomp} aggregates the model under a \textit{mismatched} assumption on local model parameters. This finding explains the restricted efficiency of \ac{aircomp} aggregation. 
    \item 
    To remedy the inadequacy of \ac{aircomp}, we propose the \ac{gue} framework, which models the aggregation target as a deterministic unknown and treats the local updates as its noisy observations. 
    Using this statistical model, we design a novel aggregation scheme that determines a \ac{ml} estimator of the aggregation target from the local parameters. We study the complexity and efficiency of the proposed scheme and compare it against the \ac{aircomp} method.
    \item 
    We validate the efficiency of \ac{gue} through extensive experiments. Numerical results depict that \ac{gue} offers substantial enhancement in learning performance 
    compared to conventional aggregation via \ac{aircomp}, especially at the low \ac{snr} regime. 
    Our investigation further demonstrates that minimizing the computation error does not necessarily lead to better learning performance. This supports our conjecture that the implicit assumption in \ac{aircomp} does not adequately capture the statistical goal of model aggregation in distributed learning. 
\end{itemize}

\subsection{Notation}
Throughout this paper, vectors and matrices are denoted by bold lower case letters (\emph{e.g.}, $\bvv$) and upper case letters (\emph{e.g.}, $\mH$), respectively. The $i$-th component of vector $\bvv$ is denoted by $\bvv[i]$. 
The integer set is shortened as $[n] = \{1,2,\hdots,n\}$. The diagonal matrix with entries $\{d_1,\hdots,d_K\}$ is denoted by $\diag\set{d_1,\hdots,d_K}$. The imaginary unit is denoted by $\mathrm{j}$, and the real and imaginary parts of $c\in\setC$ are denoted by $\Re(c),\Im(c)$ respectively. Other notations are defined in the text.


\section{System Model and Problem Formulation}


\subsection{System Model}
\subsubsection{Distributed Learning Setting}
Let $\mathcal{D}_k$ denote the local dataset at client $k$. For a sample $\zeta_i \in \mathcal{D}_k$ and a model vector $\mathbf{a}\in\setR^L$, client $k$ computes the sample loss of the model as $f(\baa , \zeta_i)$ for some loss metric $f$. The \textit{local loss} of the model at client $k$ is hence given by
\begin{equation}
    F_k(\baa) = \frac{1}{|\mathcal{D}_k|}\sum_{\zeta_i\in\mathcal{D}_k}f(\baa, \zeta_i).
\end{equation}
The clients aim to train a global model jointly over the global dataset, \emph{i.e.}, $\mathcal{D} = \cup_{k} \mathcal{D}_k$. This learning task can be cast as the distributed minimization of the \textit{global loss function,} which is defined in terms of the local loss functions as
\begin{align}\label{eq:global}
    F(\baa) = \frac{1}{|\mathcal{D}|}\sum_{k} |\mathcal{D}_k|F_k(\baa).
\end{align}

For this task, the clients invoke the following mechanism over $T$ \textit{rounds}. At the beginning of round $t\in [T]$, the clients set their models to the current global model $\baa_{t-1}$, broadcast by the server through an error-free downlink channel. Client $k$ then updates the model locally by reducing its local loss $F_k(\baa)$ through multiple iterations of a gradient-based optimizaer. Let $\baa_{k,t}$ be the latest model vector at client $k$, and the normalized local model $\bar{\baa}_{k,t}$ is obtained by
\begin{equation}
    \bar{\baa}_{k,t}[\ell] = \frac{\baa_{k,t}[\ell] - \mu_{k,t}}{\nu_{k,t}},
\end{equation}
with the sample mean $\mu_{k,t}$ and variance $\nu_{k,t}^2$ computed as
\begin{subequations}
\begin{align}
    \mu_{k,t} &= \frac{1}{L}\sum_{\ell=1}^L \baa_{k,t} [\ell],\\
    \nu_{k,t}^2 &= \frac{1}{L}\sum_{\ell=1}^L (\baa_{k,t}[\ell] - \mu_{k,t})^2.
\end{align}
\end{subequations}
The local parameters in $\bar{\baa}_{k,t}$ are then sent to the server to be \textit{aggregated} into a global model $\baa_{t}$. Note that it is also common to send the model difference $\Delta\baa_{k,t}=\baa_{k,t}-\baa_{t-1}$, and our study can be directly applied to such scenario. 
\par In the rest of this work, we focus on \emph{federated learning}, where the local parameters are aggregated proportionally to their local data size. Our proposed method and analysis can be straightforwardly applied to other distributed learning systems.



\subsubsection{Communication Setting}
The aggregation is carried out directly by analog transmissions through a wireless \ac{mac} between the $K$ clients and the server. We assume that the clients are single-antenna, while the server is equipped with an $N$-antenna array. Since the signal in the communication system contains both in-phase and quadrature components, the transmitted model parameters are converted into the \emph{complex} domain. Hence, the clients transmit their local parameters, \emph{i.e.}, $\bar{\baa}_{k,t}$, in $\lceil\frac{L}{2}\rceil$ resource blocks. We assume synchronized transmission, such that in block $i\in[\lceil\frac{L}{2}\rceil]$, with $\ell=2i-1$, the clients send their $\ell$-th and $(\ell+1)$-th parameters, \emph{i.e.}, $\bar{\baa}_{k,t}[\ell]+ \mathrm{j}\,\bar{\baa}_{k,t}[\ell+1]$, synchronously. Without loss of generality, we focus on a sample block in an arbitrary round and denote the entry sent by client $k$ as $u_k\in\setC$, \emph{i.e.}, $u_k = \bar{\baa}_{k,t}[\ell]+\mathrm{j}\,\bar{\baa}_{k,t}[\ell+1]$ for an odd index $\ell$ at round $t$. In the sequel, we drop the indices $\ell$ and $t$ for simplicity.

\par Client $k$ constructs its transmit signal from its parameter as $x_k = \beta_k u_k$ with the \textit{scaling factor} $\beta_k\in\mathbb{C}$ that satisfies $\abs{\beta_k}^2 \leq P$ for some transmit power $P>0$. It then sends $x_k$ over its uplink channel. The server receives the signal $\byy\in\setC^N$, which is given by
\begin{align}\label{eq::awgn}
    \byy = \sum_{k=1}^{K} \bh_k x_k + \bzz,
\end{align}
with $\bh_k \in \setC^N$ denoting the vector of channel coefficients from client $k$ to the server, and $\bzz\in \setC^N$ being zero-mean complex-valued additive white Gaussian noise with variance $\sigma_z^2$, \emph{i.e.}, $\bzz\sim\mac\man(\boldsymbol{0} ,  \sigma^2_z \mI_N)$. Defining $\bxx = [x_1,\hdots,x_K]^\trp$ and $\mathbf{u} = \dbc{u_1, \ldots, u_K}^\trp$, we can compactly represent the received signal as $\byy = \mH \mB \mathbf{u} + \bzz$, where $\mB = \diag\set{\beta_1,\ldots, \beta_K}$ and $\mH = \dbc{\bh_1, \ldots, \bh_K}$.  In the sequel, we refer to $\mB$ as the power matrix and $\mH$ as the channel matrix, which are assumed to be perfectly known at the server.

\par In addition to the data transmission, the pair of sample mean and sample variance $(\mu_k,\nu_k)$ is sent to the server without any transmission error, \emph{e.g.}, using orthogonal channels with error correction codes. They are used by the server in recovering a properly scaled and shifted estimate of the global model.


\subsection{Problem Formulation}

Let $\theta\in\setC$ be the \emph{ground-truth} global parameter that the server needs to estimate after each transmission block, \emph{i.e.}, $\theta = \baa_t^*[\ell]+\mathrm{j}\baa_t^*[\ell+1]$. Here, $\baa_t^*$ denotes the global model at round $t$ in the centralized scenario, where all local datasets are available at the server. The server aims to compute $\theta$ by aggregating local parameters from the received signal $\byy$. 
In general, this operation is described by an estimator $\hat{\theta}\brc{\cdot}: \setC^N \mapsto \setC$, which extracts $\theta$ from the received signal. 

To be able to rigorously define an \textit{efficient} mapping, we need to define the notion of \textit{optimal aggregation}. \begin{definition}[Optimal aggregation]\label{def::opt agg}
    The optimal aggregation from $\byy$ is defined as
    \vspace{-2mm}\begin{align}\label{eq::opt agg}
    g(\byy) = \argmin_{\hat{\theta}\brc{\cdot}: \setC^N \mapsto \setC}  \mac(\theta,\hat{\theta}(\byy)),
    \end{align}
    for some cost function $\mac$.
\end{definition}
\noindent According to Definition~\ref{def::opt agg}, the conventional aggregation problem can then be interpreted as an inference problem. This interpretation indicates that the design of the optimal aggregation depends on the statistical model of the ground-truth global parameter $\theta$.

\par In the sequel, we show that the conventional aggregation scheme based on \ac{aircomp} can fit into this formulation by considering a basic statistical representation for the ground truth. Despite its popularity, we see that \ac{aircomp} does not capture the most effective strategy for distributed learning. We hence propose an alternative model for the ground truth, through which we can develop a new over-the-air aggregation scheme. 



\section{AirComp Aggregation and Its Inadequacy}

\subsection{AirComp Aggregation}
With perfect links, the global model parameter is computed as a weighted average of local model parameters. Let $\bww$ specifies the client weights, \emph{e.g.}, by data size ratio as $\bww[k]={|\mathcal{D}_k|}/{|\mathcal{D}|}$. To recover the normalization process, the server aggregates the local models as
\begin{equation}\label{eq::agg func}   
    f(\mathbf{u})  = \bww'^\her \mathbf{u}+\bww^\her\bmu,
\end{equation}
where $\bww'[k] = \bww[k]\nu_k$ and $\bmu[k]=\mu_k$. Conventionally, this aggregation scheme is directly extended to noisy transmission, where the server aims to realize an estimate of global model parameter from its received signal $\byy$, \emph{i.e.}, setting $\theta = f(\mathbf{u})$. The aggregation in this case is interpreted as over-the-air computation of the linear function $f(\mathbf{u})$, \emph{i.e.}, analog function computation \cite{CompMAC}. This paradigm is often referred to as over-the-air aggregation \cite{Yang2020AirCompFL}.

The use of \ac{aircomp} for aggregation implicitly imposes the following (mismatched) model on the local parameters: they are modeled as \ac{iid} Gaussian vectors $\mathbf{u} \sim \mathcal{CN}(\mathbf{0},\mI_k)$,
with mean zero and unit variance~\cite{Bayesian-Aircomp}.\footnote{The zero mean and unit variance follow from the normalization of local parameters.} Since $f(\buu)$ is linear, this implies the global model parameter $\theta$ is a Gaussian random variable. The conventional solution to the computation task in this setting is given by the 
\ac{mmse} estimator, defined as
\begin{equation}\label{eq::Aircomp MMSE}
    g_\MMSE(\byy) = \argmin_{\hat{\theta}} \Ex{\abs{\theta - \hat{\theta}}^2}{p(\theta \vert \byy)} = 
    \Ex{\theta }{p(\theta \vert \byy)}.
\end{equation}
This describes a Gaussian \ac{mmse} estimation, whose solution is given by~\cite{WC-book}
\begin{equation}\label{eq::MMSE result}
    g_\MMSE(\byy) = (\mH\mB\bww')^\her(\mH\mB\mB^\her\mH^\her+\sigma^2_z\mI_N)^{-1}\byy+\bww^\her\bmu.
\end{equation}

In (\ref{eq::MMSE result}), the aggregation is computed directly from $\byy$ in terms of the channel and customized power matrix $\mB$. To design $\mB$, we evaluate the \ac{mse} between the ground-truth and the estimated parameter as
\begin{align}
&\Ex{\abs{\theta - \hat{\theta}}^2}{} = \Ex{\abs{f(\buu)  - g_\MMSE(\byy)}^2}{} \nonumber\\
    &=\bww'^\her\bww' - (\mH\mB\bww')^\her(\mH\mB\mB^\her\mH^\her+\sigma^2_z\mI_N)^{-1}\mH\mB\bww',
\end{align}
and set the power matrix $\mB$ by solving 
\begin{align}
    \max_{\mB} \quad & (\mH\mB\bww')^\her(\mH\mB\mB^\her\mH^\her+\sigma^2_z\mI_N)^{-1}\mH\mB\bww'~\label{eq::opt_b_MMSE}  \\
    \text{s.t.} \quad & |\beta_k|^2\leq P \quad \forall k\in[K] \nonumber,
\end{align}
which minimizes the expected aggregation error.

\subsection{AirComp as Optimal Aggregation}
The \ac{aircomp} aggregation approach is readily fitted to our formulation of optimal aggregation. For this scheme, the cost function $\mac$ determines the squared error between $\theta$ and its estimate, \emph{i.e.}, $\mac (\theta, \hat{\theta}) = \Ex{\abs{\theta - \hat{\theta}}^2}{p(\theta|\byy)} $.

This interpretation of \ac{aircomp} aggregation as an optimal aggregation scheme reveals its inadequacy for distributed learning. In fact, \ac{aircomp} aggregation relies on the assumption that the local parameters are drawn from \emph{statistically independent processes} and that the ground-truth is explicitly computed as $\theta = f(\mathbf{u})$. Nevertheless, from a statistical learning point of view, this assumption is not valid. In federated learning, the local model parameters are computed by backpropagation over the same model architecture with the same initial point $\baa_{t-1}$. Assuming statistically independent parameters is hence unrealistic in this case. Following this observation, we design a new optimal aggregation scheme under a more realistic statistical model for the local parameters.


\section{Global Unknown estimation}

\par   To properly capture the goal of distributed learning in our formulation, we propose a new scheme, \ac{gue}, which sets the entries of the centralized global model as the estimation target, \emph{i.e.}, $\theta = \baa^*_t[\ell]+\mathrm{j}\baa^*_t[\ell+1]$. To this end, the ground-truth global parameter $\theta$ is characterized as a \emph{deterministic unknown}. Since the local model training essentially aims to approximate such unknown, the local parameters are modeled as sample estimators of $\theta$, which compute an erroneous estimate of $\theta$ due to their limited local dataset $\mathcal{D}_k$. With this model, 
the local parameters are described by a joint distribution parameterized with $\theta$, which we denote by $p_\theta(\mathbf{u})$. 
As a result, the distribution of the received signal $\byy$ is also parameterized with $\theta$ and obtained by 
\begin{equation}
    p_\theta(\byy) = \int p_\theta(\mathbf{u}) p(\byy|\mathbf{u}) d\mathbf{u},
\end{equation}
where $p(\byy|\mathbf{u})$ describes the channel.

\par Motivated by the results of \cite{Mandt2017SGDApproxBayes}, we assume $p_\theta(\mathbf{u})$ to be a Gaussian distribution. The mean and covariance of this distribution are obtained from the postulated distribution defined later. Building on the distribution, we derive the optimal aggregation in \ac{gue} and analyze its efficiency and complexity.

\subsection{Postulated Distribution}\label{sec::postulate dist}
In \ac{gue} framework, the \ac{kl} divergence is employed as the cost function, \emph{i.e.},
\begin{equation}
    \mac(\theta,\hat{\theta}) = \int p_\theta(\mathbf{y})\log \frac{p_\theta(\mathbf{y})}{p_{\hat{\theta}}(\mathbf{y})} d\mathbf{y}.
\end{equation} 
With this cost function, the optimal aggregation describes a standard \ac{ml} estimator which estimates the ground-truth $\theta$ from $\byy$ by maximizing the likelihood $p_\theta(\mathbf{y})$. 

\par Consistent with \ac{aircomp} aggregation, we assume that the weighted average function $f(\buu)$ describe the optimal aggregation \textit{under perfect communication links}, \emph{i.e.}, when the server has access to local parameters $\buu$. Considering the cost function in \ac{gue}, this means that $f(\buu)$ is an \ac{ml} estimator of $\theta$ from the samples collected in $\buu$, \emph{i.e.},
\begin{equation}\label{eq::f is ML}
    f(\buu) = \hat{\theta}_\ML(\buu),
\end{equation}
and we use this assumption to specify the distribution $p_\theta\brc{\buu}$. 

\begin{definition}
    The postulated distribution of local parameter $\buu$ is a parameterized family $p_\theta(\buu)$, for which $f(\buu)$ serves as an \ac{ml} estimator of $\theta$.
\end{definition}

The above definition enables us to analytically specify $p_\theta(\buu)$. To this end, let $p_\theta(\buu)$ belong to a Gaussian family, \emph{i.e.}, $\buu\sim\mathcal{CN}((\theta-\bww^\her\bmu)\bvv,\mG)$ for some $\bvv\in \setC^K$ and $\mG\in\setC^{K\times K}$. The postulated distribution is then specified by $\bvv[k] = {1}/({K\bww'[k]})$ and $\mG=\diag\set{{1}/({K\bww'[1]^2}),\hdots,{1}/({K\bww'[K]^2})}$. To show this, we note that 
\begin{align}
    \nabla_{\theta} \log p_\theta(\buu) 
    &= \bww'^\her\buu - \theta+\bww^\her\bmu,
\end{align}
where the complex partial derivative operator is defined as 
\begin{equation}
   \nabla_\theta  = \frac{1}{2}\left(\derivative{\Re(\theta)}  + \mathrm{j}\derivative{\Im(\theta)} \right). 
\end{equation}
Setting this derivative to $0$, we have 
\begin{equation}
    \hat{\theta}_\ML(\buu)  = \bww'^\her\buu + \bww^\her\bmu = f(\buu),
\end{equation}
which satisfies the condition in (\ref{eq::f is ML}).
\par Note that, similar to the assumed prior in AirComp, $p_\theta(\buu)$ contains no correlation term. However, the independence observed in \ac{gue} is for a fixed $\theta$. Furthermore, it is not an assumption; it is induced by treating $f(\buu)$ as an \ac{ml} estimator under the Gaussian model. This is in contrast to AirComp, which enforces independence by design.

\subsection{Aggregation via GUE}
Considering the postulated distribution, 
the received signal is described in terms of the unknown $\theta$ by the Gaussian distribution 
    $\byy\sim\mathcal{CN}((\theta-\bww^\her\bmu)\mathbf{r},\mSigma)$,
where $\mathbf{r} = \mH\mB\bvv$ and $\mSigma = \mH\mB\mG\mB^\her\mH^\her+\sigma^2_z\mI_N$. 
The optimal aggregation via \ac{gue} is hence given by the \ac{ml} estimator of $\theta$ from $\byy$. 
Determining the derivative of log-likelihood as
\begin{align}
    \nabla_\theta \log p_\theta(\byy)  
    = \mathbf{r}^\her \mSigma^{-1}(\byy-(\theta-\bww^\her\bmu)\mathbf{r}).
\end{align}
We set it to zero to find the solution as 
\begin{equation}\label{eq::ML form}
%
\theta = 
\frac{\mathbf{r}^\her \mSigma^{-1} \byy}{\mathbf{r}^\her \mSigma^{-1} \mathbf{r} }+\bww^\her\bmu.
\end{equation}
 Hence, the optimal aggregation in the \ac{gue} framework is 
\begin{equation}\label{eq::MVU result}
    g_\ML(\byy) = \frac{(\mH\mB\bvv)^\her (\mH\mB\mG\mB^\her\mH^\her+\sigma^2_z\mI_N)^{-1} \byy}{(\mH\mB\bvv)^\her (\mH\mB\mG\mB^\her\mH^\her+\sigma^2_z\mI_N)^{-1} \mH\mB\bvv } + \bww^\her\bmu.
\end{equation}
It is worth noting that while this estimator is computed linearly from $\byy$, the linear beamforming vector is different from that of the \ac{mmse} estimator. This follows from the fact that this estimator takes a different statistical model from the one considered in \ac{aircomp}.

\subsection{Efficiency of GUE Aggregation}
It is straightforward to see that \ac{gue} determines an \textit{unbiased} estimator of the ground-truth $\theta$ with minimum variance given by the \ac{crlb}. To show this, we first note that since the \ac{gue} estimator is a linear function of $\byy$ as shown in (\ref{eq::ML form}), it is also Gaussian, with mean $\theta\mathbf{t}^\her\mathbf{r}$ and variance $\mathbf{t}^\her\mSigma\mathbf{t}$, 
where 
    $\mathbf{t} = \frac{\mSigma^{-1}\mathbf{r}}{\mathbf{r}^\her \mSigma^{-1} \mathbf{r}}$.
Simplifying, we can write
\begin{equation}
    g_\ML(\byy)\sim \mathcal{CN}(\theta,\frac{1}{\mathbf{r}^\her\mSigma^{-1}\mathbf{r}}),
\end{equation}
which concludes that 
$g_\ML(\byy)$ is an unbiased estimator. 
The estimation error is thus characterized by its variance.

We next show that $g_\ML(\byy)$ achieves the \ac{crlb}. Under the regularity conditions,
the \ac{crlb}  states that the variance of any \emph{unbiased} estimator $\hat{\theta}(\byy)$ is lower bounded by the 
reciprocal of its Fisher information, \emph{i.e.},
\begin{equation}
    \mathbb{E}_{p_\theta(\byy)}\left[\abs{\hat{\theta}(\byy) - \theta}^2\right] \geq \frac{1}{\mathcal{I}_\byy(\theta)},
\end{equation}
where the Fisher information is defined as 
\begin{equation}
        \mathcal{I}_\byy(\theta) =  
        \mathbb{E}_{p_\theta(\byy)}\left[ |\nabla_\theta \log p_\theta(\byy)|^2\right].
\end{equation}
For the \ac{gue} estimator, we have 
\begin{align}
     \mathcal{I}_\byy(\theta) 
        &= \mathbb{E}_{p_\theta(\byy)}[|\mathbf{r}^\her\mSigma^{-1}(\byy - (\theta - \bww^\her\bmu)\mathbf{r})|^2], \nonumber \\
        &= \mathbf{r}^\her \mSigma^{-1} \mathbf{r},~\label{eq::mul_Fisher}
\end{align}
whose reciprocal coincides with the variance of $g_\ML\brc{\byy}$. Since the Fisher information exists, and integration and differentiation are interchangeable, the regularity conditions are satisfied and $g_\ML\brc{\byy}$ achieves the minimum variance. 

Considering the unbiasedness of the \ac{gue} estimator, we can design the power matrix $\mB$ to minimize the variance of the estimator, 
or equivalently, maximize the Fisher information. This leads to the following power allocation policy: 
\begin{align}
    \max_{\mB} \quad & (\mH\mB\bvv)^\her (\mH\mB\mG\mB^\her\mH^\her+\sigma^2_z\mI_N)^{-1} \mH\mB\bvv \label{eq::opt_b_ML}\\
    \text{s.t.} \quad & |\beta_k|^2\leq P \quad \forall k\in[K] \nonumber.
\end{align}
Note that this power allocation policy constitutes the same type of optimization as in \ac{aircomp} aggregation, \emph{i.e.}, (\ref{eq::opt_b_MMSE}). 

\subsection{Complexity of \ac{gue} Aggregation}
The overall complexity of either \ac{aircomp} or \ac{gue} comprises two components: (i) model aggregation, and (ii) power allocation policy. The subsequent discussion compares these components for these two methods.
\par Considering the complexity of model aggregation, both methods 
compute their estimator of the global model via linear receive beamforming, \emph{i.e.}, as $\mathbf{b}^\trp\byy+\bww^\her\bmu$. The beamforming vector $\mathbf{b}$ is computed differently in these approaches; see (\ref{eq::MMSE result}) and (\ref{eq::MVU result}). 
The two methods share the same order of computation complexity for determining $\mathbf{b}$ from the channel and power matrix: both approaches require a matrix inversion of complexity $\mathcal{O}(N^3)$. 
\par With respect to power allocation, we note that the design problems in (\ref{eq::opt_b_MMSE}) and (\ref{eq::opt_b_ML}) impose the same computational complexity on the system. In fact, both problems are non-convex, and their exact solutions are intractable. 
We thus approximate the solution iteratively via \ac{slsqp}~\cite{SciPy}. In each iteration, \ac{slsqp} builds a quadratic approximation of the objective and a linear approximation of the constraints around the current optimization variable. These approximations define a quadratic program (QP) with a least-squares penalty, which is solved by a standard QP solver; see \cite[Chapter 25]{QP_solve}.

From the above discussions, we can conclude that \ac{aircomp} and \ac{gue} impose the same computational overhead on the system. The performance gains, reported in our numerical experiments, are hence obtained at no extra computation cost. 


\section{Numerical results}

\begin{figure}
    \centering
    \includegraphics[width=0.57\linewidth]{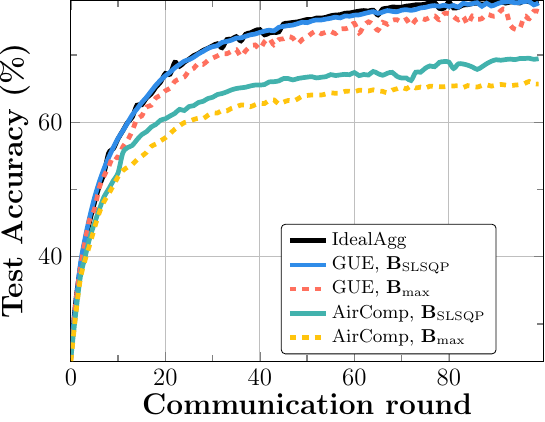}
    \caption{Test accuracy vs. round at $\text{SNR} = 5\;\text{dB}$ with $N=4$ antennas at server.}
    \label{fig::comm_round}
\end{figure}

\begin{figure*}
    \centering
    \begin{tabular}{c c c}
        \includegraphics[width=0.28\linewidth]{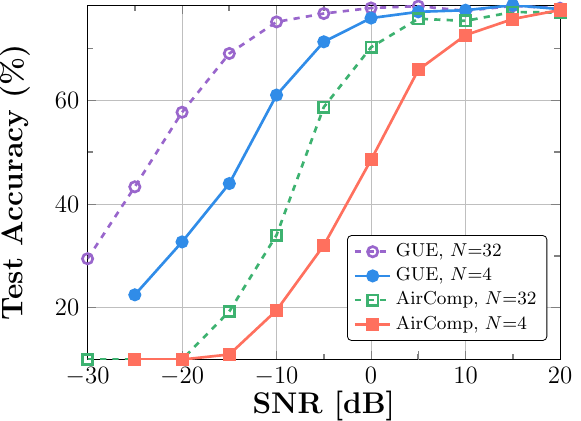}     &  \includegraphics[width=0.28\linewidth]{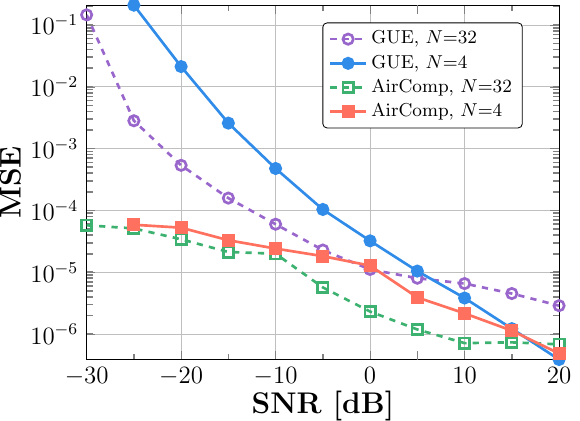} & \includegraphics[width=0.28\linewidth]{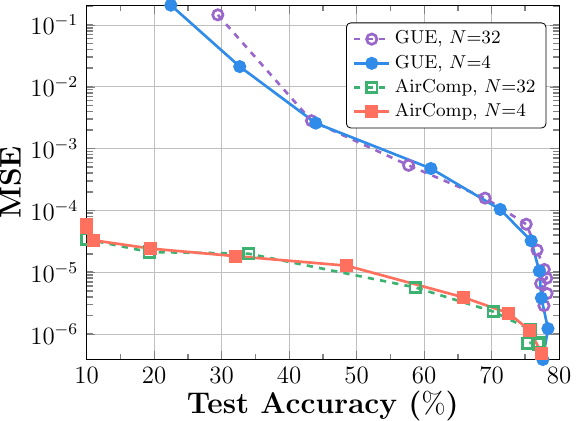}  \\
        (a) & (b) & (c)
    \end{tabular}
    \caption{Performance of GUE and AirComp aggregation with different number of server antennas $N$ : (a) and (b) show test accuracy and average \ac{mse} against SNR, respectively. (c) is showing the change of average \ac{mse} against test accuracy.}
    \label{fig::converge_performance}    
    
\end{figure*}

\par We validate our analysis through numerical experiments. 
To this end, we consider 
CIFAR-10 classification task with  
a four-layer \ac{cnn} 
with ReLU activation at hidden layers as in \cite{Seyed2024WeightedAgg}. The output is computed by a 
fully-connected layer followed by softmax. The model has roughly 
$10^5$ training parameters. The learning rate $\eta$ starts at $0.002$ and decreases following a cosine annealing scheduler. The local model of each client is trained on mini-batches of size $B=128$ over $S=30$ local iterations. 
The training loop contains $T=100$ communication rounds. 
The training dataset is uniformly distributed among $K=32$ clients. 
The channel matrix $\mH$ is full-rank and follows Rayleigh fading with channel gain $\sim\exp(0.5)$. We assume the channel coherence time is longer than the transmission duration, hence $\mH$ is fixed within each communication round. The \ac{snr} is defined as $\text{SNR} = {P}/{\sigma^2_z}$. 
For power allocation according to (\ref{eq::opt_b_MMSE}) and (\ref{eq::opt_b_ML}), respectively, for AirComp and GUE, we consider $\mB_{\text{max}} = \diag\set{\sqrt{P},\hdots,\sqrt{P}}$  for uplink transmission with maximum power, and $\mB_{\text{SLSQP}}$ when the underlying optimization is solved via \ac{slsqp}. 
In \ac{slsqp}, we initialized $\mB$ with $\mB_{\text{max}}$; thus, $\mB_{\text{SLSQP}}$ always achieves a better objective than $\mB_{\max}$.~\footnote{The code and further experimental details are available at \url{https://github.com/YQ-Comm/GUE-A-statistical-framework-for-wireless-DL}}

\par Fig.~\ref{fig::comm_round} plots the test accuracy against communication rounds for $\mB_{\text{max}}$ and $\mB_{\text{SLSQP}}$ 
at $\mathrm{SNR}=5$ dB. For comparison, the results for ideal aggregation (FedAvg~\cite{Li2020FedAvgConvergence}), \emph{i.e.}, federated averaging with perfect parameter exchange, are further shown. The results depict that at this \ac{snr} regime, \ac{gue} can closely track ideal aggregation, 
whereas using \ac{aircomp}, the training performance is significantly degraded. We also note that $\mB_\text{SLSQP}$ provides considerably better performance than $\mB_{\max}$ in both \ac{aircomp} and \ac{gue}. Therefore, in the following experiments, both methods utilize the power matrix $\mB_\text{SLSQP}$.

\par Fig.~\ref{fig::converge_performance} sketches the 
test accuracy and aggregation \ac{mse} for both \ac{aircomp} and \ac{gue} against \ac{snr} 
at different numbers of antennas $N$. Fig.~\ref{fig::converge_performance}(a) shows the maximum test accuracy, and Fig.~\ref{fig::converge_performance}(b) shows the average \ac{mse} between the true aggregation and estimated parameters, both determined over $T=100$ rounds.

From Fig.~\ref{fig::converge_performance}(a), we note that the two methods exhibit comparable accuracy at high \acp{snr}. This is intuitive, as both methods become weighted averaging, which is the optimal aggregation scheme under perfect communication links. 
At lower \acp{snr}, however, \ac{gue} significantly outperforms \ac{aircomp}. In both limited-antenna ($N=4$) and full-antenna ($N=32$) scenarios, \ac{gue} achieves an approximate $15$ dB gain over \ac{aircomp}.  


From Fig.~\ref{fig::converge_performance}(b), we note that, despite its degraded learning performance, \ac{aircomp} aggregation interestingly achieves lower \ac{mse} values than \ac{gue} almost at all \acp{snr}. 
This observation aligns with our initial conjecture that the computation error of an aggregation scheme is \textit{not} an adequate metric for evaluating learning performance. In other words, an aggregation method with a lower \ac{mse} does not necessarily lead to a higher test accuracy. This observation is depicted more explicitly in Fig.~\ref{fig::converge_performance}(c), where different schemes can achieve the same accuracy with distinct \ac{mse} values. 


\par To validate the generality of our observation, 
we present the experimental results for some other datasets in Table~\ref{tab::diff dataset}. In the low \ac{snr} regime, the test accuracy achieved by \ac{aircomp} aggregation remains degraded for all datasets, whereas \ac{gue} closely tracks ideal aggregation.

\begin{table}[h]
    \centering
    \caption{Accuracy ($\%$) after $T=100$ rounds of training with $N=4$ antennas.}
    \begin{tabular}{|c|c!{\vrule width 1pt}ccc|}
    \hline
    \rowcolor{gray!25}
       \text{Dataset} & \text{SNR} in [dB]  & \textit{IdealAgg} & \textit{GUE} & \textit{AirComp}\\
       \hline
       \textit{MNIST} & -5 & 98.78 & 98.39 & 93.51\\
       \textit{MNIST} & -15 & 98.78 & 93.21 & 87.12 \\
       \textit{Fashion MNIST} & 0 & 91.57 & 90.34 & 84.76 \\
       \textit{Fashion MNIST} & -10 & 91.57 & 79.36 & 70.84 \\
       \textit{CIFAR-10} & 5 & 78.17  & 77.12 & 63.36\\
       \textit{CIFAR-10} & -5 & 78.17  & 71.29 & 32.1\\
       \hline
    \end{tabular}
    \label{tab::diff dataset}
\end{table}

\section{Conclusions}
\par We developed a new framework for model aggregation in wireless distributed learning. 
By fitting \ac{aircomp} to this framework, 
we showed that conventional over-the-air aggregation based on \ac{aircomp} implicitly considers a mismatched assumption on the local model parameters. 
Based on this finding, we proposed an alternative aggregation scheme, called \ac{gue}, which directly estimates the global model from its noisy observation. 
Our experimental results depict significant gains, showing close-to-ideal learning performance with \ac{gue} at low \acp{snr}. The results of this study suggest that over-the-air aggregation can be effectively addressed at more realistic operating points, \emph{e.g.}, at significantly lower \acp{snr}, than what earlier studies have reported. 

\bibliographystyle{IEEEtran}
\bibliography{ref}

\begin{thebibliography}{10}
\providecommand{\url}[1]{#1}
\csname url@samestyle\endcsname
\providecommand{\newblock}{\relax}
\providecommand{\bibinfo}[2]{#2}
\providecommand{\BIBentrySTDinterwordspacing}{\spaceskip=0pt\relax}
\providecommand{\BIBentryALTinterwordstretchfactor}{4}
\providecommand{\BIBentryALTinterwordspacing}{\spaceskip=\fontdimen2\font plus
\BIBentryALTinterwordstretchfactor\fontdimen3\font minus
  \fontdimen4\font\relax}
\providecommand{\BIBforeignlanguage}[2]{{%
\expandafter\ifx\csname l@#1\endcsname\relax
\typeout{** WARNING: IEEEtran.bst: No hyphenation pattern has been}%
\typeout{** loaded for the language `#1'. Using the pattern for}%
\typeout{** the default language instead.}%
\else
\language=\csname l@#1\endcsname
\fi
#2}}
\providecommand{\BIBdecl}{\relax}
\BIBdecl

\bibitem{mcmahan2017FL}
H.~B. McMahan, E.~Moore, D.~Ramage, S.~Hampson, and B.~A. y~Arcas,
  ``Communication-efficient learning of deep networks from decentralized
  data,'' in \emph{Proc. 20th Int. Conf. Artif. Intell. Statist. (AISTATS)},
  2017, pp. 1273--1282.

\bibitem{Arbaoui2024FLSurvey}
M.~Arbaoui \emph{et~al.}, ``Federated learning survey: A multi-level taxonomy
  of {FL} research areas,'' \emph{ACM Trans. Intell. Syst. Technol.}, pp.
  1--41, 2024.

\bibitem{Li2020FedAvgConvergence}
X.~Li, K.~Huang, W.~Yang, S.~Wang, and Z.~Zhang, ``On the convergence of
  {F}ed{A}vg on {N}on-{IID} data,'' in \emph{Proc. Int. Conf. Learn. Represent.
  (ICLR)}, 2020.

\bibitem{Gafni2022FL_SPperspective}
T.~Gafni, N.~Shlezinger, K.~Cohen, Y.~C. Eldar, and H.~V. Poor, ``Federated
  learning: A signal processing perspective,'' \emph{IEEE Signal Process.
  Mag.}, vol.~39, no.~3, pp. 14--41, 2022.

\bibitem{Yang2020AirCompFL}
K.~Yang, T.~Jiang, Y.~Shi, and Z.~Ding, ``Federated learning via over-the-air
  computation,'' \emph{IEEE Trans. Wireless Commun.}, vol.~19, no.~3, pp.
  2022--2035, 2020.

\bibitem{CompMAC}
B.~Nazer and M.~Gastpar, ``Computation over multiple-access channels,''
  \emph{IEEE Trans. Inf. Theory}, vol.~53, no.~10, pp. 3498--3516, 2007.

\bibitem{FedAvg-convergence-analysis}
X.~Cao, G.~Zhu, J.~Xu, and S.~Cui, ``Transmission power control for
  over-the-air federated averaging at network edge,'' \emph{IEEE J. Sel. Areas
  Commun.}, vol.~40, no.~5, pp. 1571--1586, 2022.

\bibitem{channel-adaptive-FL}
J.~Mao, H.~Yang, P.~Qiu, J.~Liu, and A.~Yener, ``{CHARLES}:
  Channel-quality-adaptive over-the-air federated learning over wireless
  networks,'' in \emph{Proc. IEEE 23rd Int. Workshop Signal Process. Adv.
  Wireless Commun. (SPAWC)}, 2022, pp. 1--5.

\bibitem{OTA-FL-heterogenous}
T.~Sery, N.~Shlezinger, K.~Cohen, and Y.~C. Eldar, ``Over-the-air federated
  learning from heterogeneous data,'' \emph{IEEE Trans. Signal Process.},
  vol.~69, pp. 3796--3811, 2021.

\bibitem{Seyed2024WeightedAgg}
S.~M. Azimi-Abarghouyi and L.~Tassiulas, ``Over-the-air federated learning via
  weighted aggregation,'' \emph{IEEE Trans. Wireless Commun.}, vol.~23, no.~12,
  pp. 18\,240--18\,253, 2024.

\bibitem{Azimi2024Lattice}
S.~M. Azimi-Abarghouyi and L.~R. Varshney, ``Compute-update federated learning:
  A lattice coding approach,'' \emph{IEEE Trans. Signal Process.}, vol.~72, pp.
  5213--5227, 2024.

\bibitem{lower-bound-est-view}
C.-Z. Lee, L.~P. Barnes, and A.~{\"O}zg{\"u}r, ``Lower bounds for over-the-air
  statistical estimation,'' in \emph{Proc. IEEE Int. Symp. Inf. Theory (ISIT)},
  2021, pp. 2358--2363.

\bibitem{OTA-estimation}
C.-Z. Lee, L.~P. Barnes, and A.~\"Ozg\"ur, ``Over-the-air statistical
  estimation,'' \emph{IEEE J. Sel. Areas Commun.}, vol.~40, no.~2, pp.
  548--561, 2022.

\bibitem{Bayesian-Aircomp}
C.~Park, S.~Lee, and N.~Lee, ``{B}ayesian {A}ir{C}omp with sign-alignment
  precoding for wireless federated learning,'' in \emph{Proc. IEEE Global
  Commun. Conf. (GLOBECOM)}, 2021, pp. 1--6.

\bibitem{WC-book}
D.~Tse and P.~Viswanath, \emph{Fundamentals of Wireless Communication}.\hskip
  1em plus 0.5em minus 0.4em\relax Cambridge, U.K.: Cambridge University Press,
  2005.

\bibitem{Mandt2017SGDApproxBayes}
S.~Mandt, M.~D. Hoffman, and D.~M. Blei, ``Stochastic gradient descent as
  approximate {B}ayesian inference,'' \emph{J. Mach. Learn. Res.}, vol.~18, no.
  134, pp. 1--35, 2017.

\bibitem{SciPy}
P.~Virtanen \emph{et~al.}, ``Scipy 1.0: Fundamental algorithms for scientific
  computing in python,'' \emph{Nat. Methods}, vol.~17, no.~3, pp. 261--272,
  2020.

\bibitem{QP_solve}
C.~L. Lawson and R.~J. Hanson, \emph{Solving Least Squares Problems}.\hskip 1em
  plus 0.5em minus 0.4em\relax Philadelphia, PA: SIAM, 1995.

\end{thebibliography}
\end{document}